# Internal wave and turbulence observations with very high-resolution temperature sensors along the Cabauw mast

by

Hans van Haren[a], Fred. C. Bosveld[b]

[a]Royal Netherlands Institute for Sea Research (NIOZ), P.O. Box 59, 1790 AB Den Burg, the Netherlands.
[b]Royal Netherlands Meteorological Institute (KNMI), P.O. Box 201, 3730 AE De Bilt, the Netherlands.

*Corresponding author*: Hans van Haren, hans.van.haren@nioz.nl

ABSTRACT

Knowledge about the characteristics of the atmospheric boundary layer is vital for the understanding of redistribution of air and suspended contents that are particularly driven by turbulent motions. Despite many modelling studies, detailed observations are still demanded of the development of turbulent exchange under stable and unstable conditions. In this paper, we present an attempt to observationally describe atmospheric internal waves and their associated turbulent eddies in detail, under varying stable conditions. Therefore, we mounted 198 high-resolution temperature 'T'-sensors with 1-m spacing on a 200-m long cable. The instrumented cable was attached along the 213 m tall meteorological mast of Cabauw, the Netherlands, during late-summer 2017. The mast has standard meteorological equipment at extendable booms at 6 levels in height. A sonic anemometer is at 60 m above ground. The T-sensors have a time constant in air of $\tau_a \approx 3$ s and an apparent drift about 0.1°C/mo. Also due to radiation effects, short-term measurement instability is 0.05°C/h during nighttime and 0.5°C/h during daytime. These T-sensor characteristics hamper quantitative atmospheric turbulence research, due to a relatively narrow inertial subrange of maximum one order of magnitude. Nevertheless, height-time images from two contrasting nights show internal waves up to the buoyancy period of about 300 s, and shear and convective deformation of the stratification over the entire 197 m range of observations, supported by nocturnal marginally stable stratification. Moderate winds lead to 20-m tall convection across weaker stratification, weak winds to episodic <10-m tall shear instability across larger stratification.

# 1. Introduction

The dynamics of atmosphere and ocean are quite alike, with convective turbulent overturning and internal waves supported by stable vertical density stratification on the small scales, cyclones and eddies on the mesoscales and global circulation on the large scales. As the ocean is mainly stably stratified in density, internal gravity waves dominate the kinetic energy and are thus, paradoxically, crucial for generating turbulent eddies and vertical diapycnal mixing. A prominent source of such turbulent mixing is via internal wave breaking above sloping topography (Eriksen 1982; Thorpe 1987).

In the lower atmosphere of the planetary boundary layer, breaking internal gravity waves, sometimes shortened to 'gravity waves' or 'internal waves', are expected to be less important for turbulent exchange than in the ocean as daytime convection is a more dominant source of turbulence that can extend up to the tropopause at about 12 km above ground (e.g., Schenk 1974; Gettelman et al. 2002). Nevertheless, tropospheric internal waves are known to exist, as the troposphere is often stably stratified due to the action of radiation divergence. More specifically, internal waves occur during nocturnal periods in its lowest few 100 meters when radiative surface cooling stabilizes the atmosphere. Internal waves are for instance excited over mountains (Alexander et al. 2007; Akylas 2010; Sun et al. 2015; Conrick et al. 2018; La et al. 2020). Several qualitative studies using acoustic and lidar imaging have demonstrated such internal waves (e.g., Finnigan 1988; Fritts and Anderson 2003; Nappo 2002).

In this paper, we aim to detail via high-resolution observations the behavior of internal waves in the stable, nocturnal, atmospheric planetary boundary layer. The instrumentation used here has been developed for measuring internal wave-induced turbulence in the ocean, and we want to address its usefulness for measuring in the atmosphere. For this purpose, a 213 m tall meteorological mast was equipped with 198 high-resolution temperature sensors sampling at a rate of 1 Hz in late summer. Building on several previous studies (Section 2), we also seek to compare ocean internal wave - turbulence structures with those in the lower atmospheric boundary layer.

In Section 2 we present relevant parameters and observables in a theoretical background information. In Section 3 the instrumentation and data processing are described. In Section 4 results from a test-experiment are presented, before the main results in Section 5. Section 6 main conclusions are discussed and some suggestions for future work are given.

## 2. Linear and nonlinear internal waves and stratified turbulence

In unstable surface layers transport is dominated by turbulence, Monin-Obukhov scaling works reasonably well to describe vertical transport as a function of the atmospheric state variables, although requiring horizontal homogeneity and which is violated by large turbulent eddies (Andreas and Hicks 2002; Foken 2006). However, for very stable cases this formalism appears to break down (Holtslag et al. 2013). In the review paper by Sun et al. (2015) it is noted that under such conditions shear generation of turbulence is significantly suppressed. Thus, the interaction between internal gravity waves and turbulence may become important. Internal gravity waves are known to be excited by topography (e.g., Steeneveld et al. 2009). But even under relatively flat conditions these waves may emerge from smaller scale surface obstacles. Internal gravity waves may also result from the action of low-level jets at the top of the stable atmospheric boundary layer, or from the action of density currents related to processes higher up in the atmosphere or from distant topography. Sun et al. (2015) use the term "dirty waves" to indicate that in reality waves often have characteristics which are far off from ideal waves, in terms of length and constant amplitudes on which our theories are based. They also note that observations that shed light on the interaction between internal gravity waves and turbulence are scarce. Such observations and theoretical development are needed as 'turbulence in very stable conditions is poorly understood and does not categorically satisfy traditional definitions of turbulence' (Mahrt, 2014). According to Mahrt (2014), the main source of turbulence may not be at the surface, but rather may result from shear above the surface inversion. The turbulence is typically not in equilibrium with the non-turbulent motions, sometimes preventing the formation of an inertial subrange.

Internal waves are supported by the density stratification up to the buoyancy, aka Brunt-Väisälä, frequency, the natural frequency of oscillation in a stratified fluid (Groen 1948), for the atmosphere defined by (e.g., Tsuda et al. 1991),

$$N = (g/\theta \cdot \partial\theta/\partial z)^{1/2}, \tag{1}$$

where g denotes the acceleration of gravity, $\theta = T \cdot (p_0/p)^\kappa$ the potential temperature in K, T the measured absolute temperature, p the pressure and $p_0 = 10^5$ N m$^{-2}$ a reference pressure. The Poisson constant reads $\kappa = (R/c_P) \cdot (1-0.24 r_v)$, for gas constant R, specific heat at constant pressure $c_P$ and water vapor mixing ratio $r_v = q/(1-q)$ for specific humidity q (in kg kg$^{-1}$). In dry air, $r_v = 0$ and $\kappa = 0.2854$. The calculation of the buoyancy frequency, via vertically detailed temperature observations supported by general atmospheric observations as employed here, may thus indicate the possible existence of internal waves, at frequencies $\omega <$

N, and/or of turbulence induced by wave breaking, at $\omega > N$. This separation between internal waves and turbulence can be understood as turbulence scales are suppressed at time-scales > $T_N = 2\pi/N$ and internal waves do not propagate freely when their periods are < $T_N$.

Matters become complex when N varies spatiotemporally. Ideally, N is determined from averaging over all turbulence spatial and temporal scales. If these scales are unknown, N may be conservatively calculated over all internal wave scales. In the lower atmosphere near the Earth surface, nocturnal quantitative temperature and other meteorological measurements have revealed internal waves (e.g., Finnigan et al. 1984; de Baas and Driedonks 1985; Duynkerke 1991; Sun et al. 2015; Mahrt et al. 2019). Some of the cited references refer to 'Kelvin-Helmholtz (internal) wave', which is understood as a particular nonlinear wave-type.

In contrast, following Carpenter et al. (2011), we refer to linear internal waves, as non-interacting waves that are mainly driven by buoyancy forces and that can freely propagate at frequencies in the band $f \leq \omega \leq N$ (e.g., Groen 1948; LeBlond and Mysak 1978), where f denotes the inertial frequency, the vertical component of the local Coriolis frequency. As the 'background stratification' N determines the fate of the entire internal wave band, it should be calculated from averages over the inertial period of half a pendulum day in time, and over several tens of meters in the vertical for typical internal wave amplitudes in the lower atmosphere. Length scales matter, and large-scale waves may deform N via straining in locally thin strongly stratified layers separated by more homogeneous layers above and below. Such layering can be persistent in time and horizontal space to the extent that it can support small-scale, high-frequency internal waves up to small-scale buoyancy frequency $N_s > N$. The $N_s$ is thus computed over shorter vertical scales. It also reflects smaller temporal scales than those used for computing N.

Linear waves transfer momentum, but not mass or matter. In principle, one expects their spectral appearance to peak with limited bandwidth if the source has a deterministic frequency. In ocean observations, the internal wave band is spiked around tidal harmonic frequencies. However, in geophysical environments the supporting stratification varies in time and space over a broad range of scales and causes a wide distribution of energy with frequency resulting in strong 'intermittency' with time for internal waves in general. Thus, under such conditions linear waves can become nonlinear waves that will deform and may break and generate turbulent eddies whereby mass and matter are transported.

When nonlinear waves are generated via interaction with wind/current by the process of shear instability they are named Kelvin-Helmholtz instability 'KHi', or vorticity generated waves. Other forms of nonlinear waves are, e.g., solitary waves and waves generated following Holmboe instability. The KHi are an important route for transferring energy from linear internal waves to turbulence in the ocean (e.g., Smyth and Moum 2012).

KHi generation may be computed from the one-dimensional shear flow stability analysis of Miles (1961) and Howard (1961). Depending on the viewpoint, stabilizing stratification or destabilizing shear, a criterion is defined when internal waves are supported or turbulence is generated. By indicating the ratio between the regimes of stabilizing stratification and destabilizing shear, the (bulk) gradient Richardson number is defined as,

$$Ri = N^2/S^2, \qquad (2)$$

in which $S = ((\partial u/\partial z)^2 + (\partial v/\partial z)^2)^{1/2}$ denotes the vertical shear-magnitude of (wind) vector $\mathbf{W} = [u, v]$. In a one-dimensional shear flow the two regimes dominate above and below a 'critical' $Ri_{cr} = 0.25$, respectively. Following this stability analysis, KHi may develop for $Ri < 0.25$. In an essentially three-dimensional background, geophysical (nonlinear) internal wave interactions yield a transitional Ri-value of 1 between stable ($Ri > 1$) and unstable ($Ri < 1$) regimes (Abarbanel et al. 1984; Zilitinkevich et al. 2008; Freire et al. 2019). This leaves a condition of 'marginal stability' for values in the range $0.25 < Ri < 1$ (van Haren et al., 1999). Here, the viewpoint is taken from destabilizing shear generating any turbulence, even when occurring sporadically. This is different from a stability viewpoint, which allows (some) turbulence to exist in layers with local $Ri < 0.7$ that are termed 'very stable' (Sorbjan and Czerwinska, 2013). The latter can only be understood when a restratifying agent is present for smoothing the turbulent patches to retain overall stability.

In the ocean, shear is concentrated in small-scale layers where also the largest stratification is found. The dominant shear is at the inertial frequency, as internal waves at f have the smallest vertical scales whilst propagating nearly horizontally (LeBlond and Mysak 1978; van Haren et al. 1999). For the atmosphere, Finnigan (1999) also proposed a dominant shear imposed by inertial motions from above affecting small-scale layers near the boundary. In contrast with the ocean observations, Finnigan (1999) suggested that high shear layers overlapped with low stratification layers and vice versa, thereby generating a patchy intermittent occurrence of turbulence where locally $Ri < Ri_c$ in an otherwise stable boundary layer forced from above by inertial oscillations or by propagating (low-frequency) internal waves (Finnigan et al. 1984).

Duynkerke (1991) analysed standard meteorological observations sampled at once per 10 min during a summertime shallow fog night. This sampling rate resolved 'internal waves' with typical periods of 40 min, which were also observed over mountainous terrain by La et al. (2018). Distinction was observed in the lower 30 m above ground where radiation fog was found, together with stronger stratification and 'oscillatory' motions near the buoyancy frequency that were trapped in the layer up to 60 m above ground (Duynkerke 1991).

While internal waves have been mostly revealed in the nocturnal atmospheric boundary layer, turbulence intensity was not found negligible (Nieuwstadt 1984). Although Nieuwstadt (1984) focused on planetary-surface frictionally induced turbulence, an alternative suggestion may be internal wave breaking in the stable nocturnal stratification. Frehlich et al. (2008) demonstrated 'stably stratified turbulence' with dissipation rates between $10^{-5}$ and $10^{-3}$ m$^2$ s$^{-3}$ over 250 m of the atmospheric boundary layer using high-frequency profiling instrumentation and lidar.

Results from extensive arrays of 34 thermocouples and 6 sonic anemometers sampling at rates O(10) Hz the lower 60 m of the atmospheric boundary layer have been reported by Burns and Sun (2000) and Sun et al. (2012). Their analyzed data were also at 10 min intervals, gave standard deviations of 0.1°C from shielded sensors under nocturnal conditions, and yielded similar results as in Duynkerke (1991). However, for detailed quantitative studies on the transition from atmospheric internal waves to turbulence one has to sample faster for prolonged periods of time of several weeks, as has been done using instrumentation resolving the buoyancy period by a factor of 100 in the ocean (e.g., van Haren and Gostiaux 2012). The potential use of this instrumentation for the atmospheric boundary is employed and described here. Such sampling may also be feasible using distributed temperature sensing 'DTS' from optic fibre techniques in the atmosphere, suspended from a balloon (e.g., Keller et al. 2011) or attached to a mast (Peltola et al. 2021).

## 3. Experimental setting and data handling

*a. Temperature sensors*

For the specific study of internal wave turbulence in the atmospheric boundary layer we taped 198 Royal Netherlands Institute for Sea Research 'NIOZ' temperature 'T'-sensors, version NIOZ4, to a nylon-coated steel cable at 1 m intervals. NIOZ T-sensors are custom-made designed and built for independent operations in the ocean down to 6000 m (van Haren et al. 2009; van Haren 2018). The electronics are protected from the harsh deep-ocean conditions like ambient static pressure, up to $6\times10^7$ N m$^{-2}$, and salt water by corrosive-free high-grade titanium housing 0.18 m in length, 0.023 m in diameter and weighing 0.23 kg. A glass tip holds two temperature sensitive Negative Temperature Coefficient NTC resistors. The tip is about 1 mm$^3$ of the size of the Kolmogorov spatial scale. On a vertical string, typically 100 sensors are used to study internal waves and large turbulent overturning motions in stratified environment nominally sampling at a rate of 1 Hz for the duration of 1 year. This resolves the Ozmidov scales of largest energy-containing turbulent overturns in a stratified environment, but not the dissipative Kolmogorov scales.

As data transport is costly from remote deep-ocean areas, the T-sensors operate as stand-alone, running on a single Lithium AA-battery and storing data on a micro-SD-card. Lithium batteries have a flat response until a 'sudden' collapse upon draining and require bookkeeping of their use, in contrast with exponentially decreasing power level in alkaline batteries of which the remaining power can always be monitored. Internal clocks are synchronized via induction. In water, the glass-embedded NTC-Wien Bridge oscillator-based T-sensors have time constant of $\tau \approx 0.4$ s as verified following spike response on an oscilloscope, a precision of <0.0005°C and a noise level of <0.0001°C.

In air, the T-sensors' time constant is different from that in water as air has about 500-times smaller heat capacity than water. Environmental factors such as radiation effects may deteriorate the T-sensor signals. This is verified during a test experiment in Section 4. For the lower atmosphere experiment, the T-sensors also sampled at a rate of 1 Hz. Like in the ocean, this sampling rate is considerably faster than the typical nocturnal buoyancy period $T_N > 100$ s, so that large-scale turbulent eddies having time-scales between 2 and 100 s are expected to be resolved, besides internal waves. The sampling was synchronized to a single standard clock every 4 hours, so that an entire 197 m high profile was measured within an internal clock accuracy of ±0.01 s. In ocean deployments the electric circuit for the synchronization pulse through the insulated steel cable is closed via the conducting salty seawater. In air, and fresh water, the electric circuit is closed via an extra electrician's wire taped to the coated steel cable.

In the ocean, the steel cable also functions as a strength member and is tensioned tautly to at least 2 kN by sub-surface buoy(s) on top. In air, the 70 kg weighing 200 m long and 198 T-sensors holding string needs alternative support for being tightly vertical. We used a mast for that purpose.

*b. Mast set-up*

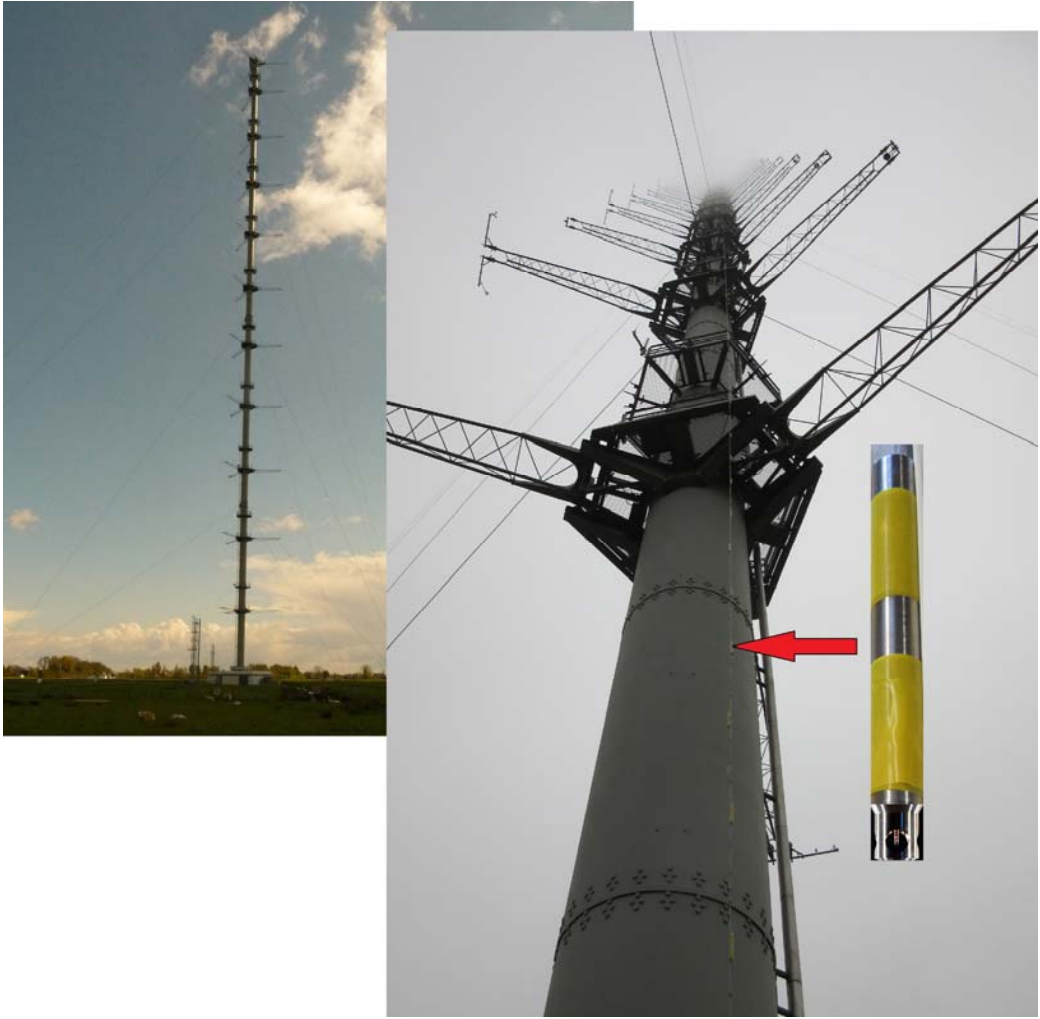

**Fig. 1**. View from the north of the entire 213-m tall KNMI-mast in Cabauw in summer, and, in the foreground, from the south and from below in autumn-fog with lower end of temperature T-sensor cable suspended along-side (the red arrow points at a 0.18 m long T-sensor).

For 50 years, the Royal Netherlands Meteorological Institute 'KNMI' operates a 213 m high guyed mast (Fig. 1) for atmospheric boundary layer research located in the flat grasslands of Cabauw, the Netherlands, at 51.971°N, 4.927°E (Bosveld et al. 2020). The mast has permanent standard and special meteorological equipment mounted at 9.4 m long extendable booms every 20 m height level. Instruments can be reached from intermediate platforms around the mast by moving the booms to their respective service level. 10-minute mean wind, temperature, and humidity observations are obtained at 2, 10, 20, 40, 80, 140, and 200 m above ground. The mast is also equipped with a Gill-R3-100 sonic anemometer/thermometer for turbulence observations at 60 m above ground. The 'sonic', as it is called in short, samples at a rate of 10 Hz and has a verified time constant of <0.1 s, so that it resolves turbulence scales down to about 0.5 m. In a field 200 m north of the mast, surface energy budget observations are performed. These energy budget observations include in- and outgoing short- and long wave radiation, and the fluxes of momentum, heat, and water vapor. An overview of previous turbulence work at Cabauw can also be found in Bosveld (2020) and Bosveld et al. (2020).

For the internal wave turbulence study we used observations from the mast equipment for the bulk background information, including the computation of N, |S|, Ri and net radiation QNET. For the detailed temperature observations, the cable with 198 NIOZ T-sensors was temporarily suspended from the 206 m service level to the roof of the Cabauw mast's main building. For safety and logistics reasons, the T-string cable was positioned on the south-side of the mast, fixed against the outside of intermediate platforms at a distance of 1.6 m from the 2 m diameter mast (Fig. 1). To minimize errors due to mast influence, we will focus on observations when the T-string was upwind from the mast. The T-string, which consisted of two sections, was led along the edge of the service and boom levels between the south guy and the south-east booms. The T-string was hoisted into the mast after lowering a pulley with hoist line from the 206 m level to the ground, pulling it up by hand and attaching the hoist line to a car. Driving the car away from the mast pulled the T-string up. One T-string-section of 99 m was placed between the 206 and 106 m service levels and another section of 100 m between the 106 m level and about 2 m above the roof of the main building. The lowest sensor was thus at $z = 7$ m, the highest at $z = 204$ m. The two sections were connected at the 106 m level. At the top of the roof, a 200 kg weight was placed and a tension band was used to secure the cable. This operation was performed on 14 August 2017 (yearday 225, using the convention that 1 January 1200 UTC = 0.5 yearday). The mounting was secured at each intermediate platform to be able to withstand a hurricane, in theory. The cable was never seen to wobble during varying winds. The T-string was lowered by hand via a pulley to the ground on 29 November 2017.

*c. Post processing T-sensor data*

Post-processing of the T-sensor data involves transfer of raw numbers to SI units using a function established from laboratory-bath calibration data. It also involves correction for the non-randomly varying bias, which may be different for every sensor. For NTCs, this bias is basically due to their electronics drift. It is unknown what precisely causes the drift, even though it is often reported in temperature metrology.

In the ocean, such a drift is noticed when measurements are made in waters with temperature variations in the 0.001°C-range showing horizontal bars in a time-depth image. Typical drifts are 0.001°C/week initially and 0.001°C/mo after aging of the NTCs. The correction for drift is made when many sensors are mounted on a cable, in an environment that eventually is inherently stably stratified. Thus, we use an intrinsic property of natural water bodies that are heated from above. In such an environment, vertical T-profiles averaged over at least the buoyancy period should be stably stratified and, when averaged over all internal wave scales including the inertial period, should be smooth as well (van Haren and Gostiaux 2012). Because such averaging includes the smoothing of turbulence at all scales, any deviations from a smooth stably stratified mean profile are then attributable to artificial drift, which is corrected for by subtracting a constant value per sensor. In practice, the averaging period varies, between the inertial period and several days, depending on the natural variation of stratification locally.

In air however, this (drift) correction is difficult for measurements that are possibly affected by radiation effects. Even after correcting over periods as short as the buoyancy period during nocturnal observations, apparent drift or measurement instability of about 0.015°C remained (see Section 4), which was impossible to remove sensibly using the above method. Applying a correction using an averaging period shorter than buoyancy period would also include (part of) genuine turbulent eddies in the average profile and is thus not appropriate. This hampers quantification of turbulence values and the computation of vertical wavenumbers from T-sensors in air. It is noted that some qualitative turbulence results may still be inferred from spectral frequency analysis which is hardly affected by the bias.

During post-processing for the present atmospheric experiment, the raw high-resolution temperature data from Cabauw mast were transferred to SI units following a laboratory calibration in a custom-made bath that can hold up to 200 T-sensors, like for ocean-experiments. The calibration is performed in water, with the sensor tips in a titanium plate. The reference thermometer and the well-insulated bath ensure calibrations with a precision of about 0.0001°C (van Haren 2018).

In air, the precision is much larger. However, this larger precision is not due to instrumental electronic noise, which is also about 0.0001°C like in water, but to poorly correctable instrumental instability, even for nocturnal statically stable profiles. As an example, the 2.4 h (0.1 d) mean profile of uncorrected 'raw' temperature data demonstrates considerable 'noise' of small-scale temperature variations with a one standard deviation spread of about 0.04°C (Fig. 2). This apparent noise is mainly short-term bias attributed to NTC-mismeasurement and a few large spikes. After correction using a high-order polynomial fit and correcting for the dry adiabatic lapse rate $\Gamma_d = -dT/dz = g/c_p = 0.0098°C\ m^{-1}$, the 'noise' of small-scale temperature variations having a one standard deviation spread of about 0.015°C in an individual 1-Hz sampled potential temperature profile appears to be smaller than the spread of wiggles in the 2.4 h mean raw data profile. Despite this reduction in small-scale temperature variations, it remains unclear what part is due to measurement errors, and what part due to genuine atmospheric motions. Inspection of the 2.4 h mean and the individual potential temperature profiles demonstrates generally stably stratified profiles with warmer air above cooler air and a largest temperature gradient near the ground. Several tens of meters high wiggles in the individual potential temperature profile around the mean profile indicate internal wave oscillations. The question is whether the <10 m high unstable wiggles demonstrate genuine turbulent eddies.

The polynomial fit to the T-sensor data provides a smooth θ-profile, which is used to calculate the mean N, As the measurement instability in air prevents turbulence quantification by splitting the signal in a reordered stable profile and turbulence displacements, the buoyancy frequency cannot be determined from the reordered profile. In order to calculate buoyancy frequency profiles from temperature profiles we minimize turbulent instabilities by averaging over 120 s and 10 m of data, using a high-order polynomial fit.

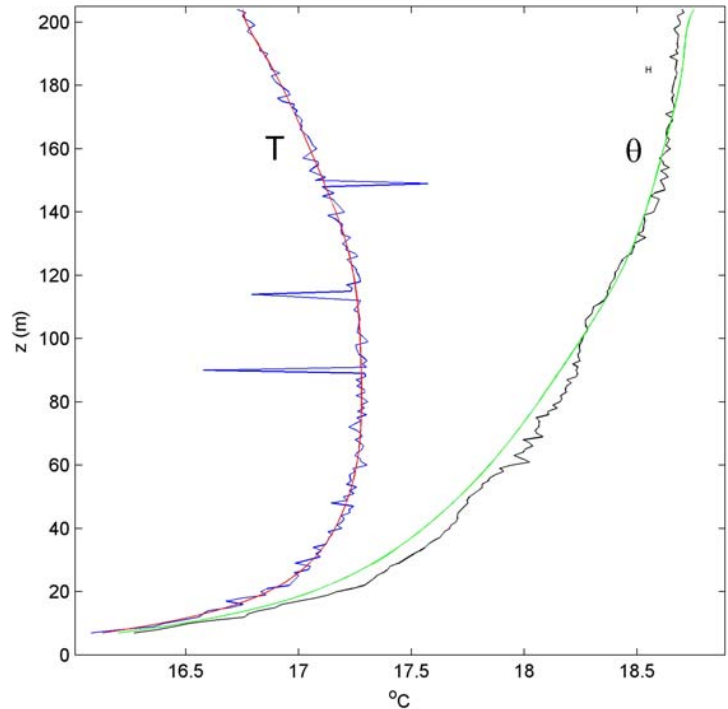

**Fig. 2**. Example of post-processing corrections for a nocturnal vertical temperature profile. In blue, the measured temperature averaged over 2.4 h between nocturnal days 228.9 and 229.0. In red, its 9$^{th}$ order polynomial fit to correct all individual profiles for T-sensors drift. This is the lowest order polynomial that provides a smooth profile, low standard error and correct continuation at the upper and lower ends. In black, a single individual profile of potential temperature observed at day 228.950 after drift- and dry adiabatic lapse rate corrections. The spread of small-scale temperature variations, 'noise' around the large-scale trend in the upper 30 m of the individual 1-Hz sampled potential temperature is 0.015°C, as indicated by the small error-bar. In green, the 2.4 h average of potential temperature profiles.

We will mainly focus on periods when the relative humidity <99% so that the air is not saturated with moisture. In this way, (dry air) potential temperature is a good tracer for density variations. Henceforth, we use the general wording of 'temperature' where we mean 'potential temperature'. For practical convenience we will use the unit °C rather than K.

## 4. Test experiment

Prior to the mounting of the 197-m long cable we performed a test experiment in the Cabauw mast. The purpose of this test was to verify the possibility of measuring air temperature fluctuations with NIOZ T-sensors, to establish the T-sensors' time constant in air, and to characterize radiation influences on the measured temperatures. Between 18 April and 5 July 2017, two T-sensors were mounted close to the sonic anemometer/thermometer at the tip of the southeast boom at $z = 60$ m and 9.4 m horizontally away from the mast (Fig. 3). The standard reference temperature sensors of the mast are at 40 and 80 m, and a reference temperature record for 60 m is constructed by linear interpolation. The high-resolution test T-sensors were mounted with their sensors oriented downward to decrease the influence of direct radiation as much as possible. They were clamped in Teflon to a small frame at the same vertical level, with the sensor-tips 0.23 m apart horizontally. As the T-sensors are taped to the cable we applied one sensor with black rubber near the sensor-head and the other with yellow tape to study possible changes in the radiation effects on the temperature measurements.

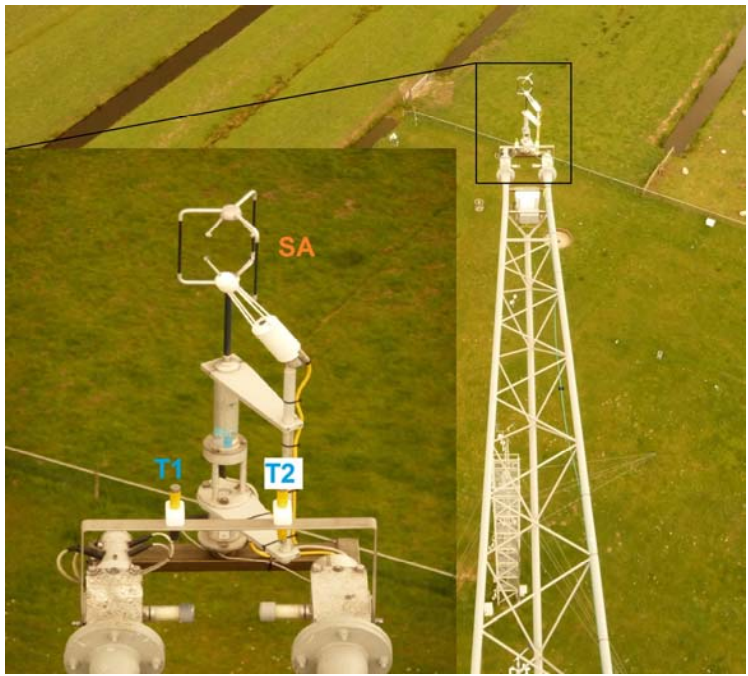

**Fig. 3**. Photos of the two downward pointing NIOZ4 T-sensors $T_1$ and $T_2$ mounted close to the sonic anemometer/thermometer SA at 60 m above ground on the tip of the SE-boom, during the test-experiment between 18 April and 5 July 2017.

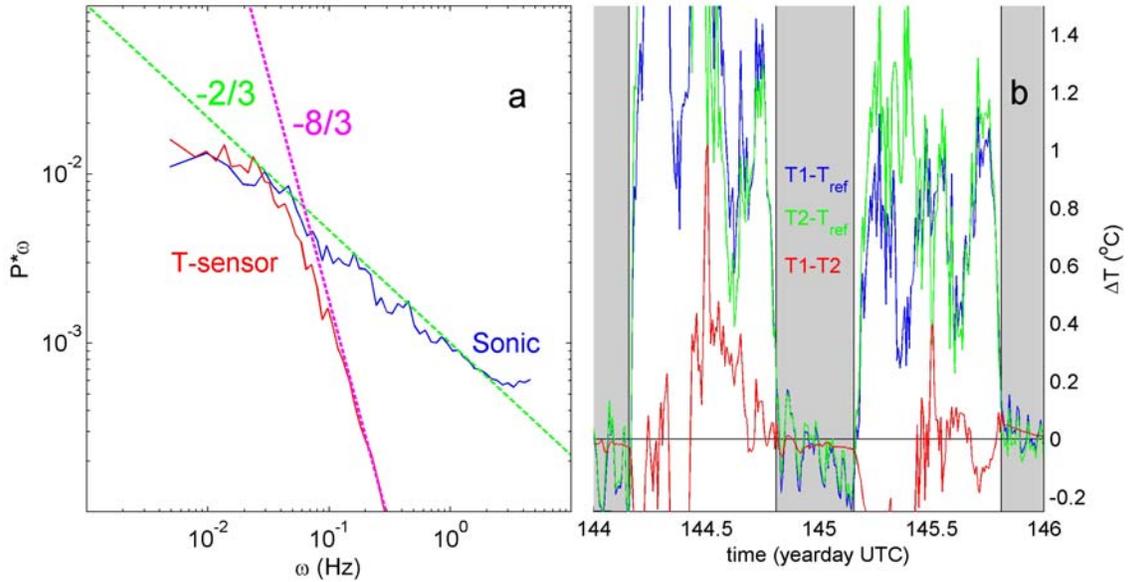

**Fig. 4**. Test experiment results. (a) 204 s mean nocturnal FFT-spectra to verify the inertial subrange ($\omega^{-5/3}$ slope; represented by the straight solid line indicating the -2/3 power-law in log-log fashion of surface area preserving spectral form, i.e. spectrum scaled by frequency) and the time constant during the T-sensor test-experiment. 10-Hz sonic data (blue) are plotted with 1-Hz T-sensor data (red). The additional straight line indicates the -8/3 power-law that represents the $\omega^{-11/3}$ slope of limited sensor time constant. The instrumental white noise level is not visible as it is very low for the T-sensors, <0.0001°C. (b) Two-day portion of 10-min mean calibrated but uncorrected data of differences between the two T-sensors (red), and the difference of the two sensors with the reference temperature (blue, green), during moderate south-westerly winds. Nocturnal periods are shaded grey.

Time constants are verified using a two-day portion of the data, when skies were clear and wind direction was such that no interference with the mast was to be expected. The first day had low wind speeds <4 m s$^{-1}$, the second day moderate wind speeds ~6 m s$^{-1}$. Spectra were computed from the original 10-Hz-sampled sonic and the 1-Hz-sampled T-sensor data. Only the high-frequency two to three orders of magnitude are considered to establish the inertial subrange of turbulence with the well-known -5/3 power law (Tennekes and Lumley 1972).

When plotted in a log-log fashion this provides a -5/3 slope, that is a -2/3 slope in an area preserving spectrum (Fig. 4a), which indeed fits the sonic data very well down to about 0.03 Hz in this example. The fit almost extends to the Nyquist frequency of the spectrum (5 Hz), after accounting for aliasing, which implies that the sonic time constant is faster than 0.1 s.

The T-sensor data also demonstrate the inertial subrange, but over half to one order of magnitude only, down to about $10^{-2}$ Hz. The low instrumental noise level <0.0001°C of the T-sensors is below the horizontal axis. At their high-frequency side they more rapidly, before 1 Hz, roll off to instrumental noise fitting to about -11/3 slope, that is -8/3 in an area preserving spectrum, due to incomplete sampling in comparison with the sonic spectrum. The two slopes (of -2/3 and -8/3) cross at about $6\times10^{-2}$ Hz, which corresponds with a time constant of $\tau \approx 3$ s for the T-sensors. This time constant is similar to optical fibre DTS (Peltola et al. 2021). No significant difference in $\tau$- and inertial subrange determination was found between the two T-sensors, between daytime and nocturnal periods and between weak and moderate wind conditions.

The relatively slow time constant of the NIOZ4 T-sensors in air implies that for typical wind speeds of 5 m s$^{-1}$ scales of 15 m are resolved. This is quite different in water, where for typical 0.2 m s$^{-1}$ current speeds length-scales of <0.1 m are resolved: Much less than the typical separation distance between the T-sensors on a string.

During nocturnal periods of the two-day data portions, little bias was found between any of the three temperature records, and a mean difference of 0.06±0.1°C was observed between the two NIOZ T-sensors and the reference sensor compared to about 0.025±0.015°C between the two NIOZ T-sensors (Fig. 4b). During daytime, however, solar irradiation caused significant over-estimation of temperature. The T-sensor temperatures showed a fair amount of scatter with typically 0.5-1°C differences in comparison with the standard reference temperature record. This scatter was presumably due to varying irradiation due to clouds, and to varying shielding against solar radiation due to the specific configuration of the T-sensors and solar position. The deviations of the T-sensors relative to the reference sensor due to radiation effects were found to correlate with wind speed, with smaller deviations at higher wind speeds. For comparison, DTS-observations by Sigmund et al. (2017) at 2 m above a meadow showed larger variations between 1 and 2.8°C due to conduction errors. After correction, the rms error between DTS and reference thermometer was 0.42°C. This is close to an rms radiation error of 0.61°C as found for an uncorrected white fiber-optic cable (de Jong et al., 2015).

The different coatings of yellow tape and black rubber did not have any noticeable effect, as the two sensors hardly showed bias between them. Over our entire 47-day test experiment, the trends in the two T-sensor records deviated by approximately of 0.15°C from each other and from the reference record, which suggests a bias of about of 0.1°C/mo. Overall over shorter time scales of one hour however, the reference temperature showed a smaller standard deviation of temperature fluctuations compared with the T-sensors during daytime periods, and a larger standard deviation during nocturnal periods. The former suggests a biased measurement of T-sensors during daytime referred to above. The latter points at a slower time constant in the T-sensors. This shows that the T-sensors are less suitable for studying daytime situations, but seem suited for nocturnal periods, when internal gravity waves are likely to occur. During nocturnal periods the T-sensor performance was best and the one standard deviation spread of noise error was about 0.015°C (Fig. 4b). As this is much larger than the instrumental noise level of 0.0001°C and varying over 10-min time scales, the T-sensor data are not suitable for turbulence quantification for which the measurement instability must be smaller and slower varying with time over at least the inertial period.

## 5. Results

A comparison between two sets of two neighbouring T-sensor records near the top and the bottom of the mast of 3 months duration confirmed the test-experiment long-term drift of about 0.1°C/mo.

Due to an error in bookkeeping of the T-sensor use, the unknown remaining power in the sensors was considerably less than expected. Already after two weeks sensors stopped and at the end of the 3 months of operation only 35% of the sensors worked. Here we consider only data from the first two weeks of operation, when at least 90% of T-sensors were operational. Missing data from 19 non-operational sensors, and removed data from noisy sensors and sensors with calibration problems, are randomly distributed along the vertical. These missing data are linearly interpolated between data from neighbouring sensors.

*a. Conditions overview*

An overview of the meteorological conditions during the first two weeks when the T-sensors were suspended in the mast is given in (Fig. 5). Wind speeds were <15 m s$^{-1}$ (Fig. 5a). For investigation of a detailed magnification we selected periods when the wind direction was from the sector SE to SW (Fig. 5b), as during the first half of Fig. 5, so that the T-string was not in the wake of the mast. During nocturnal periods, surface cooling yielded a stable potential temperature stratification of typically $\Delta\theta$ = 4-8°C over the $\Delta z$ = 197 m range (Fig. 5c). During daytime, near-homogeneous temperatures were measured, although not permanently. During the week with (south-)westerly winds from sea between days 226 and 233, a few days of relatively large specific humidity (Fig. 5d) and light rain (Fig. 5e) were registered. Otherwise, days were without precipitation. Periods with <100% relative humidity, of non-saturated air to avoid instrumental problems, occurred even for nocturnal periods (Fig. 5f). One of such nights, from day 228 to 229, with increased shear compared to daytime is selected for detailed investigation of T-sensor internal wave observations in Figs 6 to 9. A contrasting period from day 238 to 239 with weaker and easterly winds, less shear, smaller net outward longwave radiation, stronger stratification but also <100% relative humidity, except at 2 m, will be presented in Figs 10 and 11. The (nocturnal) conditions for these contrasting periods are given in Table A1.

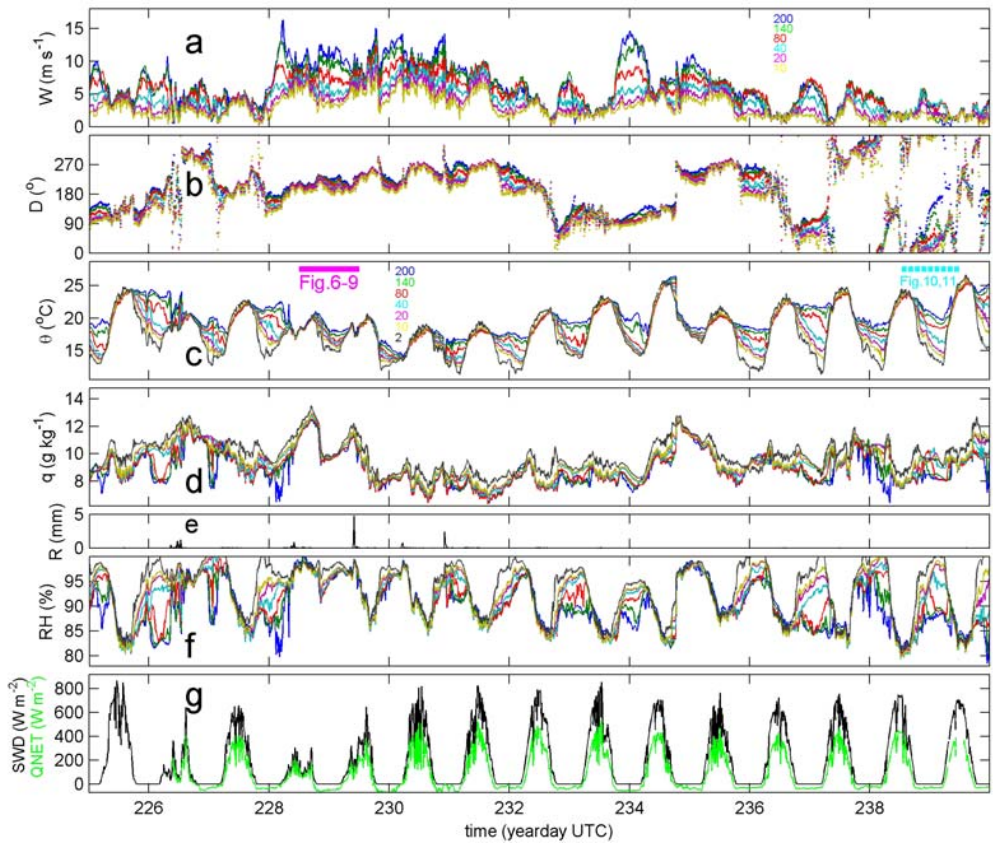

**Fig. 5**. 10-Minute sampled meteorological data overview of the first 15 days of T-string measurements. Indicated with different colours are measurements from different heights

level (in m, see legends in a. and c.). (a) Wind speed. (b) Wind direction, following the meteorological convention that 180° indicates wind coming from the South, 'southerly wind'. (c) Potential temperature (for dry air), with horizontal bars indicating one day periods and detail-figure numbers. (d) Specific humidity (aka moisture content). (e) Rain fall observed at $z = 1$ m above ground. (f) Relative humidity. (g) Downward short-wave radiation (SWD; black) and net radiation (QNET; green) observed at $z = 1$ m above ground.

*b. Detailed observations: Moderate winds, strong shear*

We selected day 228.5 to 229.5 (Fig. 6) that is characterized by moderate southerly winds with speeds <12 m s$^{-1}$ and relatively strong wind shear up to $8 \times 10^{-3}$ Hz. Wind speeds increased after sunset at high levels > 40 m and decreased at low levels (Fig. 5). Relative humidity was fairly high between 90 and 97% and light rain <1 mm occurred in the beginning until day 228.55, no rain and small humidity change during the night, and a small shower near the end at day 229.42. During the day, the downward short-wave radiation was relatively low <400 W m$^{-2}$. During the night, the mean net outward longwave radiation was relatively large: 50 W m$^{-2}$.

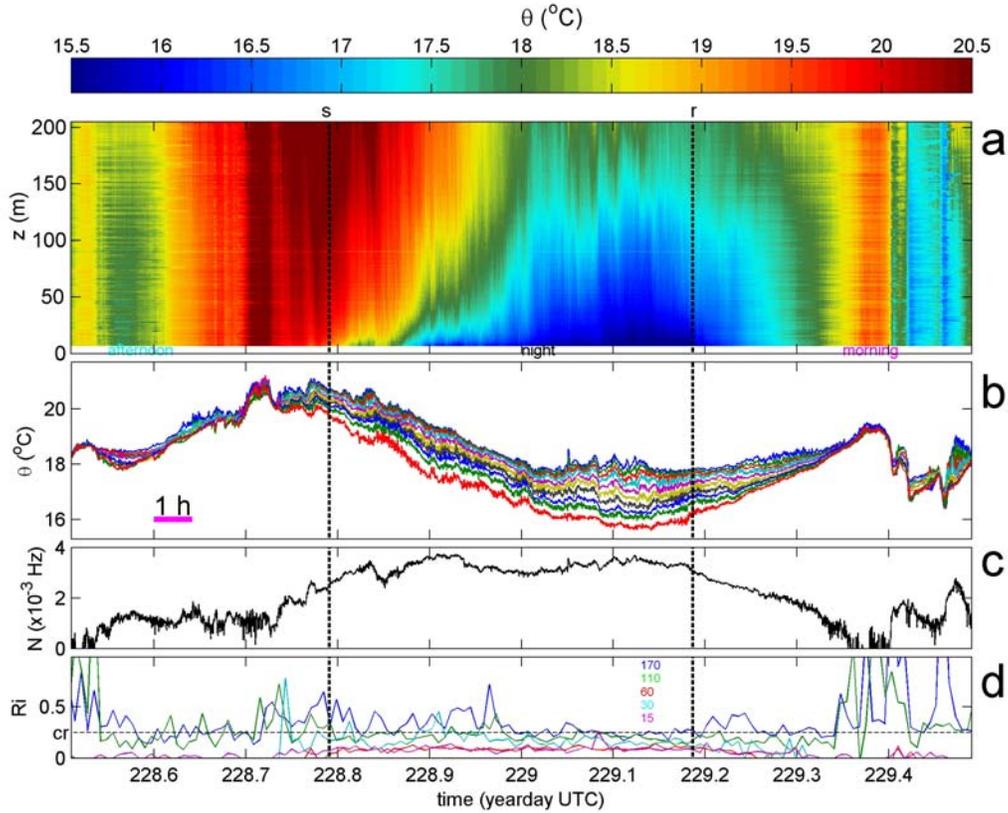

**Fig. 6**. One day example of T-string data during moderate southerly winds and little variation from 90 to 95% in relative humidity between daytime and nocturnal periods. During the nocturnal period, mean net radiation is -50 W m$^{-2}$. Local times of sunrise 'r' and sunset 's' are indicated by vertical black dashed lines. Time is in yeardays UTC, which is about 40 min behind local solar time LT. 23 T-sensors showed battery, noise or calibration problems. Their data are linearly interpolated from data of neighbouring sensors. (a) Time-depth series of potential temperature. The ground is at the horizontal axis. (b) Time series of potential temperature from height-levels every 20 m. (c) Time series of 197-m vertically averaged buoyancy frequency computed from upper and lower data in a., b. (d) Time series of gradient Richardson number Ri using (2) with atmospheric instrumentation data in Fig. 4, for five levels above ground as indicated in m. The horizontal dashed line indicates the critical level of 0.25, for one-dimensional shear flow.

In the high-resolution T-string data, neutral to convective conditions occur as $\Delta\theta < 0.5°C$ near-homogeneous temperature over the entire 197 m range (Fig. 6a, b). During daytime, besides convection, episodic weak stratification occurs of up to $N = 1.6\times10^{-3}$ Hz (Fig. 6c), under the notion that daytime T-sensor response may be biased.

About one-and-a-half hours before sunset, cooling from the ground, induced by negative net-radiation, starts a stable stratification rising upwards. Indeed the stable stratification well before sunset is governed by the surface energy balance. As evaporation is almost always very close to the reference evaporation, the sensible heat flux will become negative well before the net radiation becomes negative. This is not related to an 'oasis effect', a local microclimate that is cooler than surrounding dry area due to evaporation of a water source or plant life. However, the polders in the low lands of the western part of the Netherlands have all maintained water-level, and thus the dominant vegetation type (grass, surrounding in a wide area Cabauw mast) transpires at the reference (potential) rate (de Bruin et al., 2016). This stratification increases in strength over time higher up, and decreases with time near the ground (see below). About three hours before sunrise, temperature increases slightly at $z > 100$ m above ground, so that stratification increases as $z < 100$ m temperatures continue to decrease. About an hour before sunrise, minimum temperatures are reached at the lowest observational level. Although stratification is still considerable, this does not coincides with absolute maximum stratification computed over the 197 m range. Subsequently, the stratification decreases due to the heating from below and, also, from above, but it is not reaching the low 'daytime levels' until 3 h after sunrise at day 229.35.

During the night, the overall stratification rate yields a maximum 197-m averaged large-scale buoyancy frequency of $N \approx 3.5 \times 10^{-3}$ Hz, so that the shortest freely propagating internal waves have periods of about $T_N \approx 280$ s. Despite this apparent stability, the destabilizing increased wind shear causes Ri < 0.25 over most of the vertical during most of the night, except for the upper 50 m up to day 229.0 (Fig. 6d). Thus, if the linear stability criterion is important as bulk parameter, the nocturnal period should show distinctive turbulence levels. Overall, the low stability of the apparent stable stratification is obvious, also on smaller z,t scales.

1) FURTHER MAGNIFICATION AND VARYING STRATIFICATION

When computed over shorter vertical scales of 1 m, local buoyancy frequency varies over one order of magnitude from about $8 \times 10^{-3}$ to $8 \times 10^{-4}$ Hz between close to the ground and the highest mast levels (Fig. 7). We indicate the shortest buoyancy period in thin strongly stratified layers by the maximum 1-m small-scale buoyancy frequency $N_{s,max} \approx 10^{-2}$ Hz and the longest buoyancy period in weakly stratified layers by $N_{min} \approx 10^{-3}$ Hz. This allows small-scale 100-s internal waves to freely propagate in z < 30 m only, where $N_{s,max}$ is found, such waves will be evanescent at higher levels.

In the magnification time-depth plot of temperature, shortest 1-m-small-scale, mean 197-m-large-scale and overall longest buoyancy periods are indicated by horizontal bars, for reference (Fig. 7a). The shortest buoyancy period indicates the smallest freely propagating internal wave period possible for the image plotted. It usually occurs nearest to the ground where local stratification is largest, although decreasing in magnitude after mid-night (Fig. 7b). Amplitudes of high-frequency waves vary rapidly with time, which confirms the intermittent character of internal waves. Motions having periods shorter than this shortest buoyancy period cannot, de facto, represent freely propagating internal waves, and likely represent a form of turbulent eddies. The longest buoyancy period reflects the highest frequency internal waves that certainly propagate freely supported by the minimum stratification in the domain plotted. Between the shortest and longest buoyancy periods, motions are expected to be a mix of freely propagating internal waves and episodic turbulent overturning.

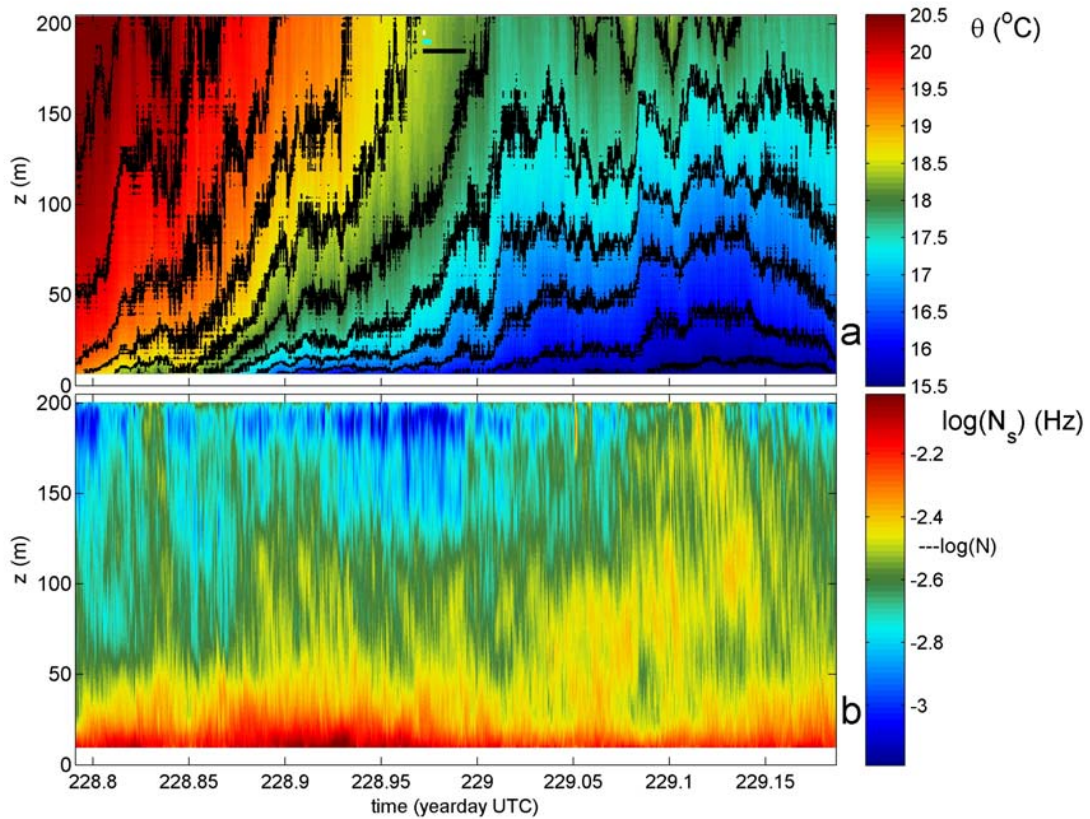

**Fig. 7**. (a) Nocturnal magnification of Fig. 6a. The white, blue and magenta bars in the upper-left corner reflect the duration of the shortest 1-m-scale, mean and overall minimum buoyancy periods of freely propagating internal waves, respectively. Black contours are drawn every 0.5°C. (b) Logarithm of small-scale buoyancy frequency computed over 1 m intervals, averaged over 120 s, and smoothed over about 10 m using a $13^{th}$ order polynomial to minimize instabilities. The value of the mean 197-m-large-scale buoyancy frequency is indicated by '- -log(N)'.

2) SPECTRA

The atmospheric temperature variance spectra are not characterized by peaks, but by broadband signals that show various slopes with frequency (Fig. 8). In the spectrum of the particular night, the log-log scale slope varies from approximately -2 at $\omega < N_{min}$, via -7/5 at $N_{min} < \omega < N$, (and possibly -1 approaching white noise 0 at $N < \omega < N_{s,max}$,), to -5/3 at $N_{s,max} < \omega < 3N_{s,max} \approx$ roll-off due to limitations of the T-sensors' time constant. It is noted that the strictly Eulerian fixed-in-space mast-observations do not provide Doppler shifted data on freely propagating internal waves, under the reasonable assumption that the wind-field does not transport the internal wave source (Gerkema et al., 2013). This will be demonstrated in the presentation of detailed observations during different wind conditions in Section 5c.

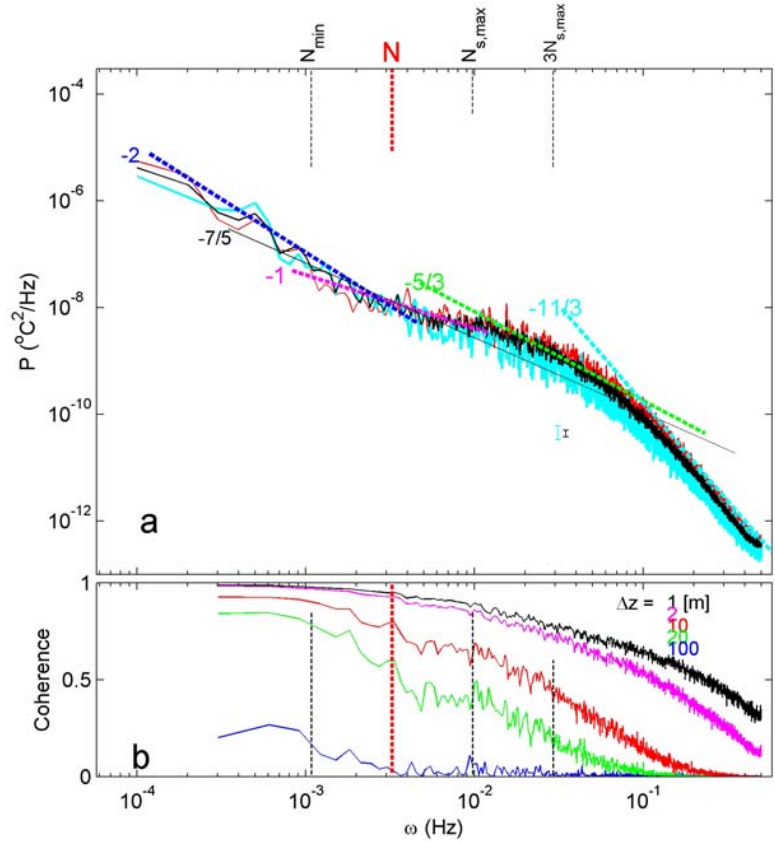

**Fig. 8**. Unscaled temperature spectra, averaged over several ranges of T-sensors, for the nocturnal period between sunset and sunrise of Fig. 6. (a) Power spectra of internal wave band and turbulence subranges for all T-sensors (black) and for those from the upper 50 m (light-blue) and lower 50 m (red) of the vertical range. Different spectral slopes (for a log-log plot) and buoyancy frequencies $N_{min}$, $N$, $N_{s,max}$ are indicated, see text for explanation. (b) Corresponding coherence between all possible pairs of independent T-sensors at different mutual distances as indicated. The 95% significance level is approximately at coherence = 0.1.

In oceanography, a -2-slope reflects the canonical ocean-interior saturated internal wave slope for frequencies $f \ll \omega \ll N$ (Garrett and Munk 1972). Recall that freely propagating internal waves are mostly expected in the frequency range $f \leq \omega \leq N$ (f is to the outside left of Fig. 8a). The -7/5-slope, or the Bolgiano-Obukhov scaling, is known for an active scalar convective turbulence (e.g., Pawar and Arakiri 2016). Originally, this scaling was proposed for the buoyancy range of stratified turbulence, between the internal wave band and the inertial subrange of turbulence (Bolgiano 1959). The -1-slope is equivalent to unsaturated, intermittent internal waves and has been observed close to N in the open-ocean (van Haren and Gostiaux 2009). The inertial subrange suggests a dominance of shear-induced turbulence for a passive scalar (Tennekes and Lumley 1972; Warhaft 2000). Shear production is thus expected at $\omega < N_{s,max}$, either via internal wave shear including near-inertial wave shear, or via frictional boundary layer shear. The clearest change in spectral slope is seen at N, with a local hump around $N_{s,max}$. For $\omega > N_{s,max}$, the spectrum shows slightly more variance near the boundary than higher up in the T-sensor range. This may be interpreted as a, barely significant, shift of the inertial subrange to higher frequencies near the boundary where z-scales are reduced. However, it is noted that a -5/3 slope is not well distinguishable and, instead, the shift more concerns the transfer from -1 slope to -7/5 slope at $N_{s,max}$, near the boundary where stratification is concentrated (Fig. 7b).

For the nocturnal internal wave range at $\omega < \times 10^{-3}$ Hz $\approx N_{min}$, coherence (Fig. 8b) is found high with levels >0.8 between sensors at mutual distances of $\Delta z \leq 20$ m, while for $N \approx 3 \times 10^{-3} < \omega < 10^{-2}$ Hz $\approx N_{s,max}$ such high levels are only found for the shortest $\Delta z \leq 2$ m. The former suggests local low-vertical-mode internal waves. The latter describes short-scale internal waves possibly supported by thin layer stratification. A small sub-peak is observed around $N_{s,max}$ for intermediate and larger scales, with further drop in coherence at higher frequencies. At $\omega > 10^{-2}$ Hz, the coherence levels roll off slowly, and they do not reach insignificance levels for the smallest vertical scales $\Delta z \leq 2$ m at the present Nyquist frequency. While this high-frequency end is far beyond the maximum freely propagating internal wave frequency, and not well resolved by the relatively slow T-sensors, its non-zero coherence may suggest that full 3D isotropic turbulence is not yet reached at frequencies of about 0.5 H during the nocturnal period.

3) FURTHER NOCTURNAL MAGNIFICATIONS

The spectral observations are supported by further time-depth magnifications of the T-sensor data (Fig. 9). The nocturnal magnifications are dominated by apparent internal waves that are almost uniform over the 197-m vertical (Fig. 9a,c), having typically O(10) m 'amplitudes' that vary throughout the domain and being largest for z > 50 m.

Further magnifications (Fig. 9b,d) demonstrate quasi-oscillatory motions that have periods shorter than the shortest buoyancy period (the white bar, which is longer than the 60 s of Fig. 9d). Such motions have a strongly nonlinear appearance, with an apparent downward 'phase' propagation as they are slanted in the z,t domain at a rate of about 200 m/30 s. Although limited by the T-sensor resolution, regular small overturns (closed contours) are observed, and a likely suggestion for a KHi of shear-induced overturning in Fig. 9d. Quantification of the turbulent mixing is impossible, because of the T-sensors' measurement instability correction problems. The observed KHi does not form a roll-up billow, but merely a half-turn before collapsing. This is similar to ocean observations (van Haren and Gostiaux 2010).

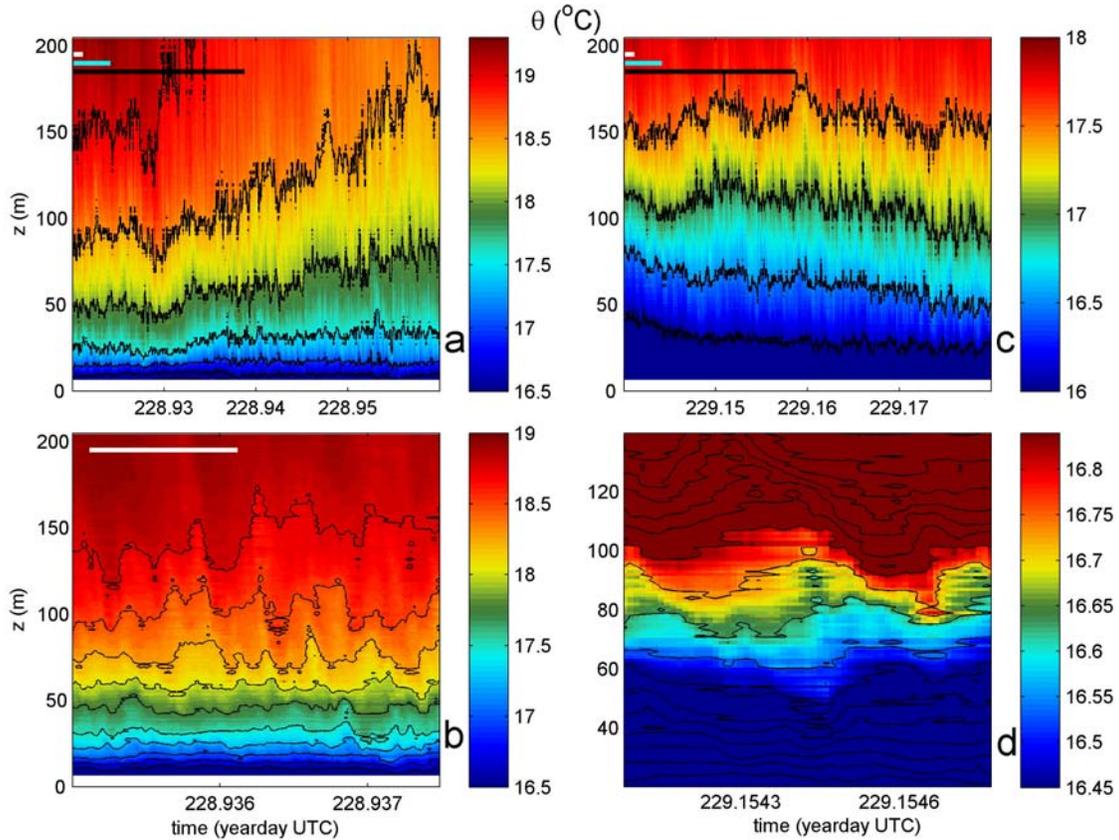

**Fig. 9**. Magnifications of nocturnal Fig. 7a with different colour ranges and black equidistant T-contours at different intervals. When the magnification window exceeds one or more buoyancy periods, the horizontal bars reflecting the periods are omitted. (a) One hour of data during 3 to 4 h after sunset. (b) A 215 s magnification of a. showing apparent wave motions that have 'periods' smaller than the shortest small-scale buoyancy period (about 85 s; white bar) and a downward 'phase' propagation. (c) One hour of data during 1.25 to 0.3 h before sunrise with multiple high-frequency internal oscillations. (d) A 60 s magnification of c., indicating an apparent local vertical mode-2 internal oscillation with strong suggestion of a shear-induced KHi of about 40 s duration and >20 m high.

*c. Detailed observations on a day with weak wind and shear*

The above one-day of late-summer observations was typical, but variations did occur between different days. Figs 10, 11, 12a show (details of) a day with strong short-wave radiation, with twice larger nocturnal stratification ($\sim N^2$), less nocturnal net outward longwave radiation, and weak north-easterly winds with little vertical shear.

Despite the different conditions, the observations in Fig. 10 show largely the same tendency as in Fig. 6. Exceptions are smaller high-frequency motions during nocturnal stratification in the time series of Fig. 10b compared to Fig. 6b and larger Ri ($> Ri_{cr}$) for $z > 20$ m in Fig. 10d compared to Fig. 6d. The one-third larger nocturnal buoyancy frequency in Fig. 10c occurs during two-thirds of the value of net outward longwave radiation associated with Fig. 6c. During the nocturnal period, the T-spectrum of Fig. 11 is smoother than in Fig. 8. It has very similar -2 slope and, even slightly larger, variance content at $\omega < N$, thus thereby extended over the range [$N_{min}$, N] in which -1 slope prevails and demonstrating no effects of the different background advecting winds. At higher frequencies $\omega > N$ variance is smaller than in Fig. 8. A -7/5 slope is found at $N < \omega < N_{s,max}$. Especially for the upper range, the -5/3 slope better fits than the -7/5-slope for $\omega > N_{s,max}$ in contrast with Fig. 8. Like in Fig. 8 albeit at smaller variance in Fig. 11, the lower and upper range spectra deviate for $\omega > N_{s,max}$, again with a small hump at $N_{s,max}$ in the lower range spectrum only where highest stratification is found. Coherence is generally smaller in Fig. 11b than in Fig. 8b at all vertical intervals and frequencies, which suggests more relative dominance of higher-vertical-mode internal waves and smaller-scale vertical overturns.

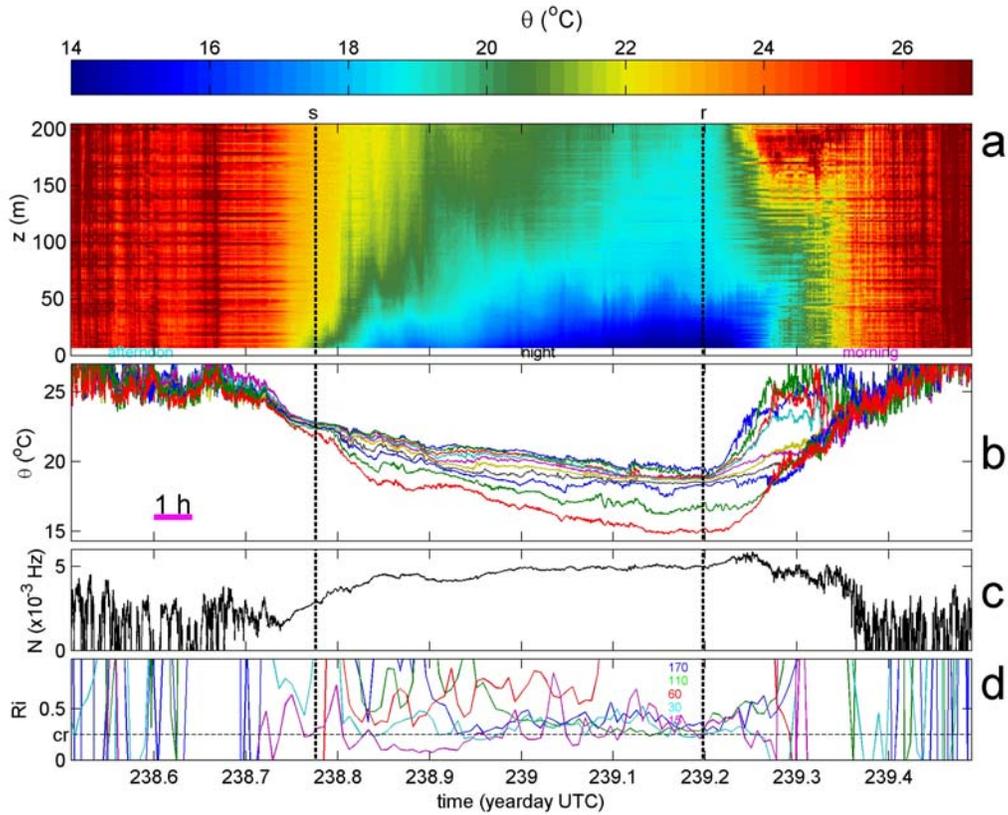

**Fig. 10**. As Fig. 6, but for a period between days 238.5 and 239.5 without rain, with strong daytime downward radiation, twice larger nocturnal stratification, 65% lower net radiation value of -32 W m$^{-2}$, weak north-easterly winds that are little sheared and less relative humidity except at ground level where it reaches 100% during late night and visibility briefly is reduced to 500-1000 m. 38 T-sensors showed battery, noise or calibration problems. Their data are linearly interpolated from data of neighbouring sensors. Note the different scales in all panels except d. compared with Fig. 6.

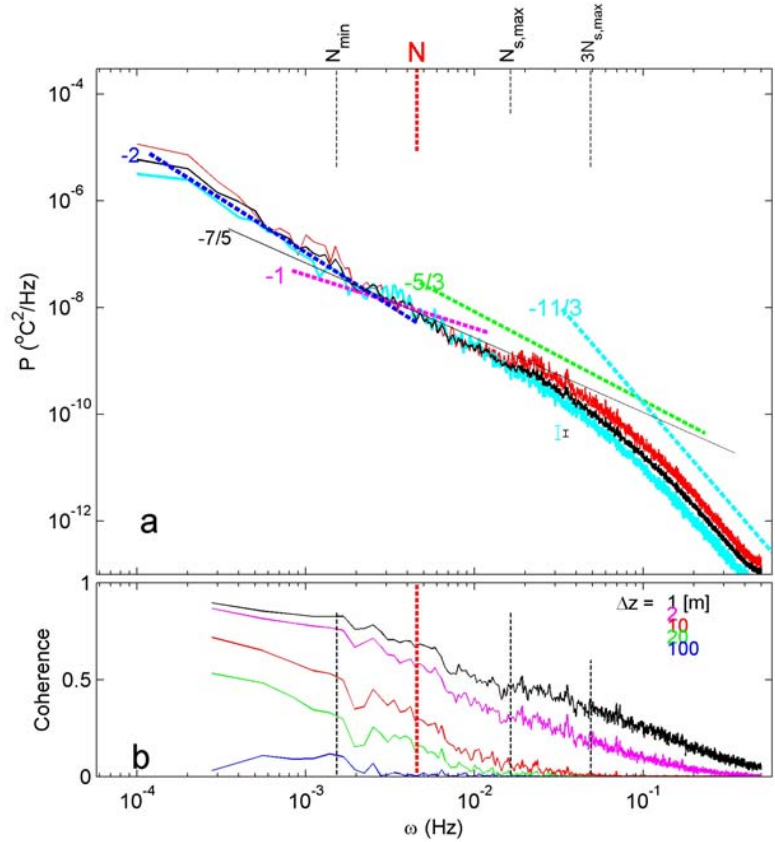

**Fig. 11**. As Fig. 8, but for nocturnal data in Fig. 10. The slopes are kept the same, for reference.

The spectral observations including the larger temperature variance, larger vertical coherence, and less steep, more convective turbulence slopes for $\omega > 6\times10^{-3}$ Hz during nocturnal period between days [228.8, 229.2] compared to [238.8, 239,2] are all visible in a further magnification comparison (Fig. 12). During the first 1.5 hours immediately after sunset, very high frequency ($\omega \gg N$) quasi-turbulent motions are not observed in the depth-time domain of Fig. 12a, contrary to the period 10 days earlier (Fig. 12b). Stratification rises more quickly with time in Fig. 12a than in Fig. 12b and the internal waves and turbulence instabilities have a linear appearance with variable 'periods' $>T_N$ in Fig.12a.

Resuming in terms of bulk parameters (cf. Table A1), the nocturnal period between days [228.8, 229.2] (Fig. 8, 12b) is characterized by more net outward longwave radiation, less stratification, more wind speed and shear, lower gradient Richardson number, less temperature variance for $\omega < N$ whilst higher temperature variance for $\omega > N$ especially near the boundary, higher vertical coherence on all scales and frequencies, and more convective than shear-induced turbulence compared with the nocturnal period between days [238.8, 239.2] (Fig. 11, 12a). These observations are confirmed during the nights immediately after the days mentioned when bulk conditions like monotonic decrease in temperature and largest stratification at lowest levels were the same.

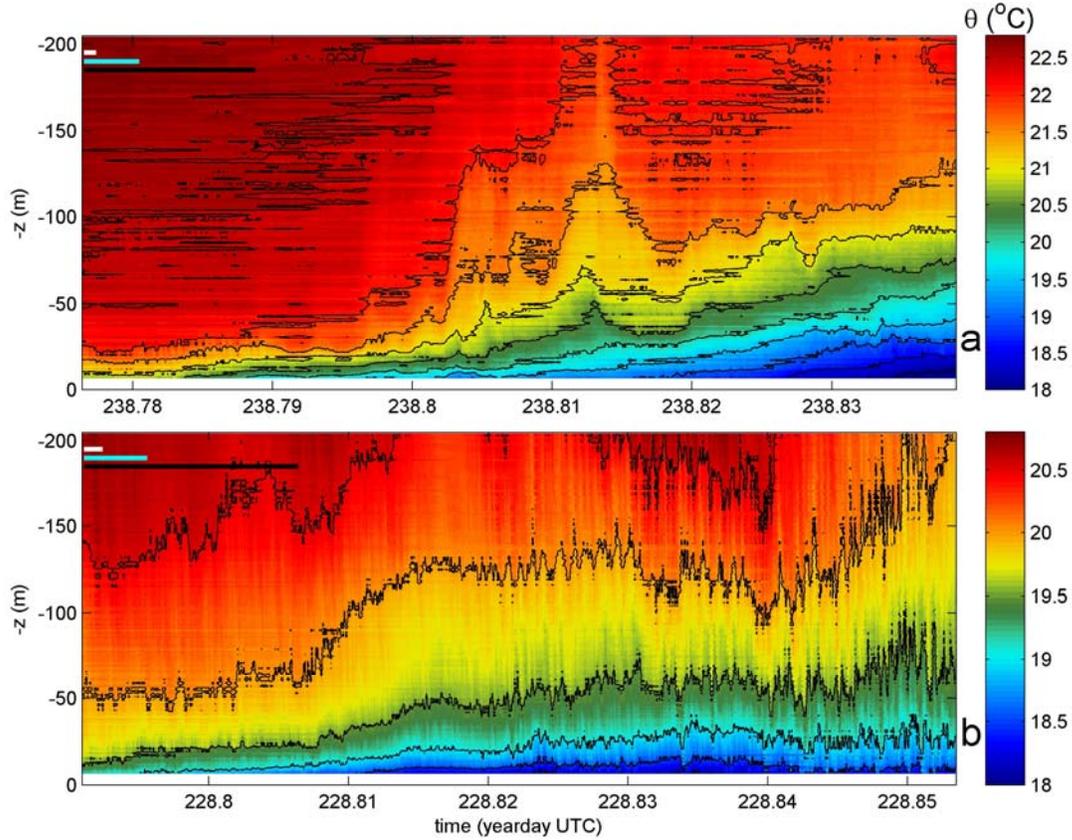

**Fig. 12**. Time-depth potential temperature details during 1.5 h immediately after sunset. In black, contours are given every 0.5°C. (a) From Fig. 10a when Ri > 0.25 for most of the vertical except for z > 150 m. (b) From Fig. 6a when Ri < 0.25 except for z < 20 m, with different colour scale.

# 6. Discussion

Previous 10-min sampled meteorological observations have revealed the existence of internal waves in the stably stratified nocturnal atmospheric boundary layer (e.g., de Baas and Driedonks 1985; Nappo 2002; Mahrt et al. 2019). Such waves were considered to be generated for Ri < 0.25 in the lower levels above ground z < 60 m where the buoyancy period was smaller than 100 s (Duynkerke 1991). It is surprising that observed much slower varying 40-min quasi-periodic motions could be trapped in this layer. Possibly, the near-ground shear essentially triggered unstable KHi that may have forced linear internal waves higher up in the troposphere, at frequencies well below N.

Like previous observations (e.g., Chimonas et al. 1999; Fritts et al. 2003; Peltola et al. 2021) the present high-resolution observations demonstrate that quasi-periodic motions occur episodically throughout the nocturnal lower atmosphere $0 < z < 200$ m. The combined analysis of high-resolution z,t imagery and spectra provides evidence of internal waves and turbulent motions. Like in ocean high-resolution temperature observations (e.g., van Haren and Gostiaux, 2012), the small-scale maximum buoyancy frequency is an important parameter to distinguish internal waves and turbulent eddies. For frequencies higher than $N_{s,max}$, the temperature variance spectrum from < 25 m from the boundary exceeds that of higher up, which implies more short-scale motions near the boundary. At these super-buoyancy frequencies, the spectra slope like -5/3 denoting an inertial subrange or -7/5 of convective overturning, depending on the conditions. At lower frequencies, the classic internal wave slope and, around large-scale buoyancy frequency N, partially intermittent partially unsaturated internal slope is observed, like in the open ocean (van Haren and Gostiaux, 2009).

The comparison between two different stably stratified nocturnal periods of moderate and weak wind speeds and shear demonstrates distinctly different internal wave turbulence that is not directly associated with frictional boundary turbulence. Although information is lacking from altitudes $z > 200$ m and potential forcing from above, the local bulk parameter observations give some insight in the different conditions. During moderately sheared winds and unstable gradient Richardson number one would expect a large inertial subrange reflecting the shear production. Whilst we do observe larger temperature variance at super-buoyancy frequencies, the associated spectral slopes are more reminiscent of convective and intermittent turbulence, especially close to the boundary.

In contrast with the ocean, the large-scale wind/current-shear is the main driver of turbulent instabilities in the apparently stably stratified nocturnal atmospheric boundary layer. For areas with a flat surface like around Cabauw mast, Finnigan (1999) proposed a flux of KHi wave energy from above causing patches sub-critical Richardson number and turbulence in the boundary layer.

During the night between days 238 and 239 end of August with low wind-shear, the above very high-frequency >$N_{s,max}$ motions reflecting convection and/or KHi are few and linear internal waves dominate the observations. The classic inertial subrange stretches over a wider range [N, $N_{s,max}$] in the upper range mainly, showing less temperature variance than under moderate winds. Also under these conditions however, 0.25 < Ri < 1 suggests marginal stability from a destabilization viewpoint, implying episodic vertical diapycnal turbulent mixing occurring (Abarbanel et al. 1984; Zilitinkevich et al. 2008), as found in the stratified North Sea (van Haren et al. 1999). This may be contrasted with the stability function following Monin-Obukhov theory, resulting in potentially very stable nocturnal stratification (Mahrt et al. 2014). Although we cannot verify the stability function as we do not have vertical velocity fluctuation measurements with the high-resolution temperature observations, the particular period with largest stratification showed relatively low net outward longwave radiation, besides weak winds, in contrast with the reasoning in Mahrt et al. (2014). The wind speed and shear are of importance. When high, they generate turbulent heat transport from the atmosphere so that the Earth surface remains relatively warm. This results in relatively low temperature gradient in the atmosphere and a relatively large upward longwave radiation and thus large (negative) net radiation. When low, the sensible heat flux is low resulting in a cooling of the Earth surface and a lower (negative) net radiation while the atmospheric temperature is larger. Unfortunately, the present high-resolution T-sensors demonstrate too large measurement instability in air to be able to quantify the vertical turbulent exchange. The problem is the relatively large time constant compared to the time scales of motions of interest. Furthermore, the time constant is not a fixed number, but slightly varies between T-sensors, thereby affecting the consistency of images and preventing quantification of turbulent mixing. For comparison, in water the time constant is 10 times smaller while, in the deep ocean, the buoyancy periods are 10 times larger resulting in a 100 times better performance than in air.

Nonetheless, the 1-Hz sampled 200 m vertical range does reveal abundant internal wave and stratified turbulence activity in the lower atmosphere, with large >100 m vertical coherent scales for low internal wave frequencies < $10^{-3}$ Hz ≈ $N_{min}$, and small <10 m scales for frequencies between $4\times10^{-3}$ Hz ≈ N and $1.2\times10^{-2}$ Hz ≈ $N_{s,max}$. At higher frequencies, quasi-periodic motions appear as nonlinear waves, with turbulent eddies on the fringes that fill the inertial subrange of shear-dominated turbulence. Even convection seems to occur in quasi-periodic episodes of 10-60 s and coherent over 20 m vertical scales also during nocturnal periods, under moderate winds mainly.

For future investigations it is suggested to find a means to improve the temperature observations. While the instrumental noise level of <0.015°C during nocturnal periods seems adequate and standard deviations of 0.05 to 0.1°C between neighbouring T-sensors are comparable to those reported by Burns and Sun (2000), the T-sensors should be made more stable to avoid measurement instability. As this instability is a combination of radiation effects, especially daytime short-wave radiation and nighttime cloud-cover varying long-wave radiation with a negligible effect of the mast, and the slowness of the T-sensors' response, improvements should be directed to these effects. Perhaps the T-sensor tip should be made smaller or of a different material, perhaps sampling the Wien bridge faster. Measurements would benefit from a faster response sensor with $\tau < 0.1$ s in air. If successful, it is expected that observations will also improve during the much stronger radiative daytime, possibly with the sensor in a small effective radiation shield.

Observational studies on the transition from large-scale shear flow, via (sheared) internal waves to instabilities leading to turbulence do exist (e.g., Finnigan et al. 1984; Viana et al. 2010; Sorbjan and Czerwinska 2013; Mahrt et al. 2019) but are still scarce (Mahrt 2014), whether above a flat grassy field or in mountainous regions. Improved instrumentation will lead to more insight, qualitatively and quantitatively, that may also help to better parameterize large-scale atmospheric modelling. A distinction that may be made in such modelling is the variation in internal-wave-turbulence in the nocturnal boundary layer as a function of varying winds and net outward longwave radiation. The present observations demonstrate the ubiquitous occurrence of internal waves of stable variance and spectral slope during nocturnal periods and their varying transition to turbulence in the lower atmosphere. The density-stratified nocturnal periods show episodic low stability via wind-shear that lead to convective overturning at relatively large vertical scales near the boundary. Marginal stability with little convection and smaller vertical scales near the boundary is found during nights when wind-shear is weak.

New observational techniques such as DTS using optical fibre technology are promising, although it is recommended for atmospheric boundary layer studies on internal waves and stratified turbulence not to suspend the light-weight instrumentation from a balloon that moves under moderate wind speeds (Fritz et al. 2021) but from a rigid mast. A short recent experiment (Peltola et al. 2021) from a 125 m mast above a forest showed DTS results with resolution (0.5 m), sampling rate (2 s), time constant (1 to 3 s) and noise levels (0.2°C) roughly similar to the present observations, albeit their evaluation was in 30 min data blocks. Longer duration experiments with DTS may prove useful for detailed investigations. For example, spectra may be computed like using the present T-sensor observations, noting that a smaller and more robust time constant is desirable. Longer duration DTS may be well supplemented with higher resolution but shorter duration turbulence techniques such as sonic and profilers like TLS and Lidar that provide profiles over 15 min and 250 m vertically with a vertical resolution of 0.8 and 15 m, respectively (Frehlich et al. 2008).

## 7. Conclusions

The 1-Hz and 1-m sampled T-sensors over a 200 m vertical range are capable of detailing internal wave and stratified turbulence activity in the lower atmosphere. Although the characteristics of the instrumentation, with a time constant of 3 s and a measurement instability of 0.015°C during nighttime, hamper quantification of turbulence, internal waves up to the 300-s buoyancy period are supported by nocturnal marginal stability. At higher frequencies, quasi-periodic motions appear as nonlinear waves, with turbulent overturning on the fringes that fill the inertial subrange of shear-dominated turbulence.

The comparison between two different stably stratified nocturnal periods of moderate and weak wind speeds and shear demonstrates distinctly different internal wave turbulence that is not directly associated with frictional boundary turbulence. During moderately sheared winds, relatively large outward radiation and unstable gradient Richardson number, the associated spectral suggest dominant 20-m tall convective turbulence, especially close to the boundary. During weak winds relatively low outward radiation, shear-induced turbulence is found dominant.

**Table A1.** Conditions for two nocturnal periods.

| Yearday | 228.8-229.2 | 238.8-239.2 |
| --- | --- | --- |
| Figure # | 6-9, 12b | 10, 11, 12a |
| W (m s$^{-1}$) | <12 | <6 |
| D (°) | 180-200 | 0-180 |
| S (Hz) | <8×10$^{-3}$ | <3×10$^{-3}$ |
| N (Hz) | <3.5×10$^{-3}$ | <5×10$^{-3}$ |
| Ri | 0.1-0.5 | 0.2-1 |
| RH (%) | 90-97 | 85-99 |
| R (mm) | <1 | 0 |
| QNET (W m$^{-2}$) | >-50 | >-32 |

*Acknowledgements.*

We thank A. Driever for the indispensable assistance in mounting and dismounting the T-sensor cable into the Cabauw-mast. The T-string was prepared with help from NIOZ-NMF. NIOZ temperature sensors have been financed in part by NWO, the Netherlands Organization for Scientific Research. M. Stastna (Univ. Waterloo, Canada) provided the darkjet colourmap suited for our data.

*Data Availability Statement.*

Requests for data can be directed to the corresponding author.